\documentclass[multphys,vecphys]{svmult}

\usepackage{makeidx}         % allows index generation
\usepackage{graphicx}        % standard LaTeX graphics tool
\usepackage{multicol}        % used for the two-column index
\usepackage[bottom]{footmisc}% places footnotes at page bottom
\makeindex             % used for the subject index
\begin{document}

\title*{Altitude Effect in \v{C}erenkov Light Flashes of
Low Energy Gamma-ray-induced Atmospheric Showers}
% Use \titlerunning{Short Title} for an abbreviated version of
% your contribution title if the original one is too long

\titlerunning{Altitude Effect in the \v{C}erenkov Light Flashes}

\author{A. Konopelko\inst{1,2}}
% Use \authorrunning{Short Title} for an abbreviated version of
% your contribution title if the original one is too long
\institute{Max-Planck-Institut f\"{u}r Kernphysik, Heidelberg
\texttt{alexander.konopelko@mpi-hd.mpg.de} \and
Humboldt-Universit\"{a}t zu Berlin, Institut f\"{u}r Physik,
Adlershof \texttt{konopelk@physik.hu-berlin.de}}
%
% Use the package "url.sty" to avoid
% problems with special characters
% used in your e-mail or web address
%

\maketitle

\begin{abstract}
At present the ground-based Very High Energy  (VHE) gamma-ray
astronomy is racing to complete construction of a number of modern
gamma-ray detectors, i.e. CANGAROO III, MAGIC, H.E.S.S., and
VERITAS. They should be fully operational in a year's time. After
much debate, the further development of this gamma-ray astronomy in
the foreseeable future must be widely anticipated to proceed with
the designing and building of a new instrumentation, which is
primarily intended for the further drastic reduction of the energy
threshold in gamma-ray observations down to about 10 GeV. On the
ground one can hardly reach such low energy thresholds without
considerably larger, up to 30 m diameter, optical telescopes, which
might be able to collect sufficient amount of \v{C}erenkov light
from the atmospheric gamma-ray showers of that low energy. If not
taken off the ground entirely (like GLAST), then it seems to be
profitable to mount future low energy \v{C}erenkov telescopes at
higher altitudes in the atmosphere in order that they will be able
to detect substantially more unabsorbed \v{C}erenkov light from a
shower. There are a few sites up on the high mountains of roughly 5
km height worldwide, which can be used for such a venture. However,
one has to remember that actual time profile, and in particular a
two-dimensional distribution (image) of \v{C}erenkov light flash
from atmospheric gamma-ray showers, undergo a rapid change after an
increase in the observational level. This paper briefly describes
the results of a topological analysis of \v{C}erenkov light images
calculated at both conventional and desirable altitudes of 2.2 and 5
km above the sea, respectively. A discussion on major upgrades of
image topology at high altitude is also given.
\end{abstract}

\section{Introduction}
The usual way to proceed with design studies for a future project,
at least in a field of VHE gamma-ray astronomy, is to perform full
scale simulations of detector response for various anticipated event
types. Apparently, gamma-ray-induced atmospheric showers represent a
sample of signal events, whereas cosmic-ray showers form a
dominating background (for review of \v{C}erenkov imaging technique
see \cite{tw04}). For a future low-threshold instrument a correct
tuning of detector design to optimize its response to gamma-ray
events becomes a most important issue, due to the fact that the
sensitivity of such a detector will be given by {\it angular and
energy resolution for signal events}. Both of these will finally
determine an efficiency of background rejection. Here we have
studied how a single parameter of detector design, particularly the
height of the observational site, may affect a topology of signal
events. Therefore, we have calculated the response of a ground-based
\v{C}erenkov light telescope of 30~m diameter, assuming an ideal
focal-plane detector. Comparative analysis of simulated events
helped us to understand what are the major differences in
parametrization of the \v{C}erenkov light flashes from gamma-ray
induced air showers, registered at two observational levels, i.e. 2
and 5 km above sea level. Considering the analysis results, there is
a discussion at the end of this paper as to which observational
level might be considered as more favorable for effective shower
imaging.

\section{Simulations}
Numerous comparisons of a few shower simulation codes
%available in
%the H.E.S.S. collaboration,
have been recently performed by different groups.
%the Monte Carlo group,
They have revealed a rather good level of agreement in basic
parameters of \v{C}erenkov light emission in gamma-ray-induced air
showers over broad energy range, starting from 100~GeV and expanding
up to 20~TeV. Calculations presented here have been carried out
using one of those simulation codes, namely ALTAI code \cite{altai}.
This numerical code was extensively used for production of the
simulated data for the HEGRA system of five imaging atmospheric
\v{C}erenkov telescopes at La Palma \cite{ko99}.

Shower simulations have been done for the standard continental
atmosphere (US standard atmosphere) \cite{elterman} for the
wavelength range of \v{C}erenkov light photons from 300 to 600~nm.
Absorption of \v{C}erenkov light photon in the atmosphere due to
Rayleigh and Mie scattering was modelled according to the data given
in \cite{elterman,valley}. The detector simulation procedure used
here accounted for all efficiencies involved in the process of the
\v{C}erenkov light propagation and registration \cite{ko99}. It
includes (i) mirror reflectivity; (ii) the acceptance of the funnels
placed in front of the photomultipliers (PMTs) (iii) the
photon-to-photoelectron conversion inside the PMTs (bi-alkali
photocathode).

Shower simulations were made here in the so-called "batch" mode. A
shower propagating time, corresponding atmospheric depth, and a
number of emitted \v{C}erenkov photons were saved for each
multiple-scattering segment of all electron trajectories in a
shower. The actual segment size was chosen as small as 0.1~[$\rm
gr/cm^2$]. CPU time needed for simulation of such low energy
gamma-ray shower ($E_o$=10~GeV), using customary computers, is
short, and it is not an issue for any scheme's optimization. One
record for a single multiple-scattering segment was treated as an
individual "batch" of emitted \v{C}erenkov photons. At the second
step of this simulation procedure all recorded batches were restored
and finally used in the calculation of the response of a number of
\v{C}erenkov telescopes, situated at different atmospheric
altitudes. Such an approach provides an opportunity to use exactly
the same simulated showers for various telescope arrangements at
different observational levels. It should be noted that the
estimated statistical error for the parameters of the \v{C}erenkov
light emission given below is $\leq 10$\%.

For the next generation of ground based \v{C}erenkov telescopes
%H.E.S.S. Phase II telescopes,
a dish of roughly 30 m diameter is foreseen.
%\cite{wh03}.
Issues around the design of such a big reflector are addressed in
\cite{wh01}. The simulation setup here consisted of 12 such
telescopes, which were arranged along one line at distances from 0
to 300 m from the shower axis. Note that the same showers were used
in calculations for each of these telescopes. It allows for a direct
comparison of the telescope's responses at different distances from
the shower axis, and the accurate study of fluctuations in
\v{C}erenkov light flashes at different shower impacts.

\section{Results}

Distribution of \v{C}erenkov light emission from atmospheric
showers can be characterized using a smooth function
\begin{equation}
\eta=\eta(t,\vec{r},\vec{\theta}), \label{aaa}
\end{equation}
which gives a mean number of photons per unit area arriving at the
observational level at time {\it t}, with space, $\vec{r}=\{x,y\}$,
and angular, $\vec{\theta}=\{\theta_x,\theta_y\}$, coordinates,
calculated with respect to the shower axis. The presentation
(\ref{aaa}) presumes averaging over a number of photons at any local
spot, because Monte Carlo simulations have provided the list of
individual photons with their coordinates. In the ideal case a
space-angular distribution of \v{C}erenkov photons in the
observational plane is simply a sum of $\delta$-functions
constructed for each individual photon.

The lateral distribution of \v{C}erenkov photons at the
observational level, which is supposed to be perpendicular to the
shower axis, is given by
\begin{equation}
\rho(r) = \rho(|\vec{r}|)= \int_0^\infty \int_{2 \pi} \eta(t,
\vec{r}, \vec{\theta} ) dt ~d\Omega.
\end{equation}
The function $\rho(\vec{r})$ is the density of \v{C}erenkov light
(the number photons hitting the unit square at $\vec{r}$).

In a similar way one can derive a temporal distribution of a
\v{C}erenkov light pulse, and a two-dimensional angular
distribution (image) of a \v{C}erenkov light flash
\begin{equation}
p(t) = \int_{A(\vec{r}_o)} \int_{\Omega_o} \eta(t, \vec{r},
\vec{\theta} ) rdr ~d\Omega_o,
\end{equation}
\begin{equation}
q(\vec{\theta})=\int_{A(\vec{r}_o)} \int_o^\infty  \eta(t,
\vec{r}, \vec{\theta} ) rdr~ dt,
\end{equation}
where $A(\vec{r}_o)$ and $\Omega_o$ are the area of the reflector
and the angular camera size, respectively, of the telescope placed
at $\vec{r}=\vec{r}_o$.

The function $\eta(t,\vec{r},\vec{\theta})$ can be well described
by a set of functions $\rho(r)$, $p(t)$, and $q(\vec{\theta})$,
given for a number of telescope locations,
$\vec{r}^{(i)}_o,~i=1,n$.

\subsection{\v{C}erenkov light density}
\label{ss:1}

\begin{figure}[t]
\centering
\includegraphics[height=7cm]{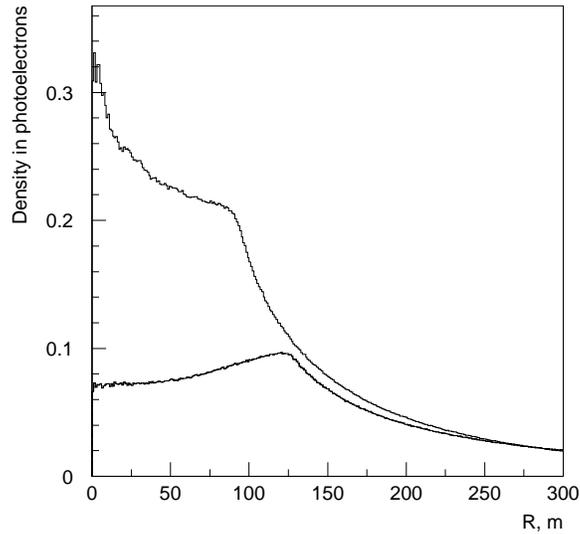}
\caption{Density of \v{C}erenkov light, $\rho(r)~[\rm ph.-e./m^2]$,
in a 10~GeV gamma-ray-induced atmospheric shower at two
observational levels of 2 and 5 km above sea level, respectively.
The density was measured in photoelectrons $\rm [ph.-e./m^2]$. A
photon-to-photoelectron conversion efficiency of $\chi= 0.1$ was
assumed here, $\rho~ \rm [ph.-e./m^2]=\chi\cdot \rho~ \rm
[photon/m^2]$. \label{fig:1}}
\end{figure}

It was emphasized in \cite{ap01}, that a substantial increase in
\v{C}erenkov light density at high altitudes in the atmosphere might
be very promising for a further reduction of the effective energy
threshold of a telescope array, which can be erected at a height of
about 5~km above the sea. One can see in Figure~\ref{fig:1} that for
a 10~GeV gamma-ray-induced atmospheric shower the density of
\v{C}erenkov light at 5~km above sea level, in a range of distances
of the telescope to the shower core limited by $r \rm \leq~100~m$,
is about a factor of 2.5 higher than the corresponding density at
2~km altitude. At the same time {\it in a range of relatively large
impact distances, $r \rm \ge 125$~m, the \v{C}erenkov light density
remains the same at both observational heights.} (see
Figure~\ref{fig:1}).

The atmospheric depth of the shower maximum can be estimated as
$X_{max}=t_o ln(E_o/E_c)$, where $t_o$ is the radiation length in
air ($t_o \simeq \rm 37~g/cm^2$), $E_o$ is the primary energy of
air-shower, and $E_c$ is the so-called critical energy ($E_c \simeq
\rm 80$~MeV). Thus for a 10~GeV $\gamma$-ray-induced air-shower the
atmospheric depth of its shower maximum is about 180~$\rm g/cm^2$. A
substantial fraction of \v{C}erenkov light photons emitted from the
shower maximum will be absorbed while traveling down to the
observational level. Between 30\% and 16\% (in the wavelength range
of 300-600 nm) for the heights of 2.2 and 5~km above sea level,
respectively. At the same time the \v{C}erenkov light pool shrinks
significantly at higher altitudes. The approximate radial size of
the light pool at 2.2~km is about 130~m (see Figure~\ref{fig:1}),
whereas at 5~km it might be roughly limited by $\simeq 90$~m. It
results in a corresponding increase of \v{C}erenkov light density by
approximately a factor of 2. This geometrical effect has a major
contribution on the increase of \v{C}erenkov light density at high
altitude as shown in Figure~\ref{fig:1}.

For a large telescope of a 30~m diameter, a requirement of a minimum
number of 15~ph.-.e.\footnote[1]{This requirement of a minimum
number of ph.-e. for the telescope trigger is in fact not a generic
value, and it might be slightly different in certain circumstances,
but it does not affect the general discussion given here.} in the
\v{C}erenkov light flash, which is sufficient to trigger the
telescope, will limit the allowed range of impact distances for
10~GeV gamma-ray showers to roughly $\rm r \leq 300~m$ (see
Figure~\ref{fig:1}). It corresponds to the same effective detection
area of $\rm S=3 \times 10^5~m^2$ at both observational heights. On
the other side, assuming that the \v{C}erenkov light density scales
with primary shower energy as $\rho\propto E^{1.1}$, one can roughly
estimate the minimal primary energy of the gamma-ray shower, which
has still sufficient amount of light at the density profile plateau
(r$<$125~m) and can still trigger the telescope. Simple calculations
yield, accordingly, a factor of 4 and 10 for 2 and 5~km altitudes,
respectively. It means that one can catch gamma-ray events of energy
$\geq 2.5$~GeV and $\geq 1$~GeV at 2 and 5~km observational height,
respectively, using the same telescope. It is important to mention
that all these extremely low-energy events will be concentrated
within a radius of roughly 100~m, which is determined by the actual
shape profile of a lateral distribution function of \v{C}erenkov
light. The drastic drop in photon density beyond 120~m will prevent
the detection of such gamma-ray showers at larger distances to the
shower axis. As a result the detection area for these events will be
a weak function of primary shower energy. Furthermore, the detection
area of high energy gamma-ray showers will be of the same size at
both observational heights.

\begin{table}[t]
\centering \caption{Parameters of the fit in Eq.~\ref{eqn:1}. D
[$\rm gr/cm^2$] denotes an atmospheric depth at a given
observational level. R is the impact distance of the telescope to
the shower axis. \label{tab:1}}
\begin{tabular}{lllllll}
\hline\noalign{\smallskip}
H [km] & D [$\rm gr/cm^2$] & R [m] & C & $\alpha$ & $t_o$ & $\beta$  \\
\noalign{\smallskip}\hline\noalign{\smallskip}
5 &500 & 50  & 2.595$\times 10^{-5}$  & 8.533  & 3.463 & 13.778\\
&    & 100 & 2.650$\times 10^{-10}$ & 14.149 & 4.670 & 19.076\\
&    & 150 & 2.424$\times 10^{-11}$ & 12.495 & 6.500 & 15.827\\
&    & 200 & 4.824$\times 10^{-13}$ & 12.154 & 9.085 & 15.081\\
&    & 250 & 7.352$\times 10^{-14}$ & 11.277 & 12.240 & 13.770\\
    \hline
2.2 &843 & 50  & 1.360$\times 10^{-6}$  & 7.689 & 5.824 & 19.295\\
&    & 100 & 2.871$\times 10^{-15}$ & 17.936 & 6.620 & 32.634\\
&    & 150 & 4.031$\times 10^{-17}$ & 17.907 & 8.157 & 26.966\\
&    & 200 & 1.225$\times 10^{-18}$ & 17.551 & 9.988 & 23.094\\
&    & 250 & 3.512$\times 10^{-20}$ & 17.264 & 12.368 & 21.476\\
\noalign{\smallskip}\hline
\end{tabular}
\end{table}

\subsection{Time profile of \v{C}erenkov light flash}
The longitudinal extension of an atmospheric shower, and its
location in space with respect to the telescope, finally determine a
time profile of the \v{C}erenkov light flash. For a gamma-ray shower
of a certain primary energy, e.g. of 10~GeV, recorded at fixed
observational level (e.g. 2 or 5~km above sea level), the shape of
the \v{C}erenkov light flash basically depends only on the actual
distance of the telescope to the shower axis. The arrival time of
\v{C}erenkov photons onto reflector is $t=t_e+t_{\check{C}}$, where
$t_e$ and $t_{\check{C}}$ is accordingly a propagation time of
emitting electrons and photons, respectively. Electrons of an
energy, which is sufficient for emission of \v{C}erenkov light in
the atmosphere, are apparently moving faster than the emitted
photons. Thus at relatively small distances of the telescope to the
shower axis ($r\leq$120~m) \v{C}erenkov photons, emitted at later
stages of the shower development in the atmosphere, are actually
arriving earlier than the photons emitted at the beginning of shower
development. It swaps over at a large distance from the shower axis,
because of the rather long travel path in dense atmosphere for the
photons emitted by electrons out of a dying particle
cascade\footnote[2]{This is a well-known effect, sometimes called
the "sea-gull" effect, in the shape of \v{C}erenkov light pulses
(e.g. see \cite{kruger})}. Therefore, in general, the larger the
distance of the telescope to the shower axis, the broader the
corresponding \v{C}erenkov light pulse.

Time pulses of \v{C}erenkov light flashes always show a very steep
rising edge, and a slow fall-off. They can be well fitted by
\begin{equation}
p(t)=C\cdot t^{\alpha}(1+(t/t_o)^{\beta}). \label{eqn:1}
\end{equation}
Parameters of the fits of time pulses simulated for two
observational levels and for a number of impact distances are
given in Table~\ref{tab:1}. The normalization condition was
$\int_o^{50ns}p(t)dt=1$. The contribution of \v{C}erenkov photons
delayed by longer times than 50~ns is negligible.

\begin{figure}[t]
\centering
\includegraphics[height=7cm]{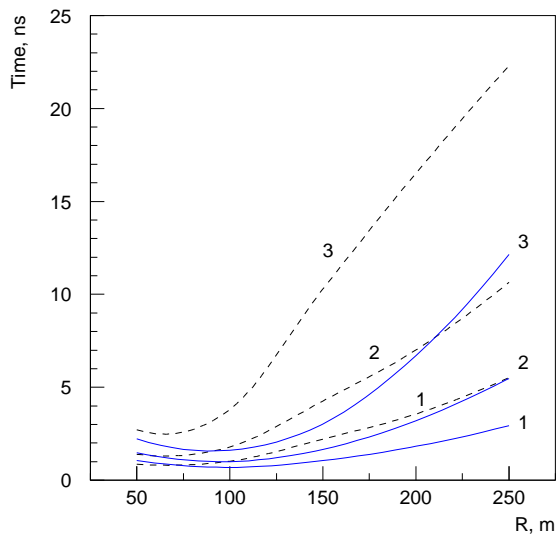}
\caption{Parameters of \v{C}erenkov light time pulses, $t_1$ (1),
$t_2$ (2), $t_3$ (3), calculated at two observational levels of 2
(solid curves) and 5~km (dashed curves) above sea level, as a
function of impact distance of the telescope to the shower axis R.
\label{fig:2}}
\end{figure}

\begin{figure}[t]
\centering
\includegraphics[height=7cm]{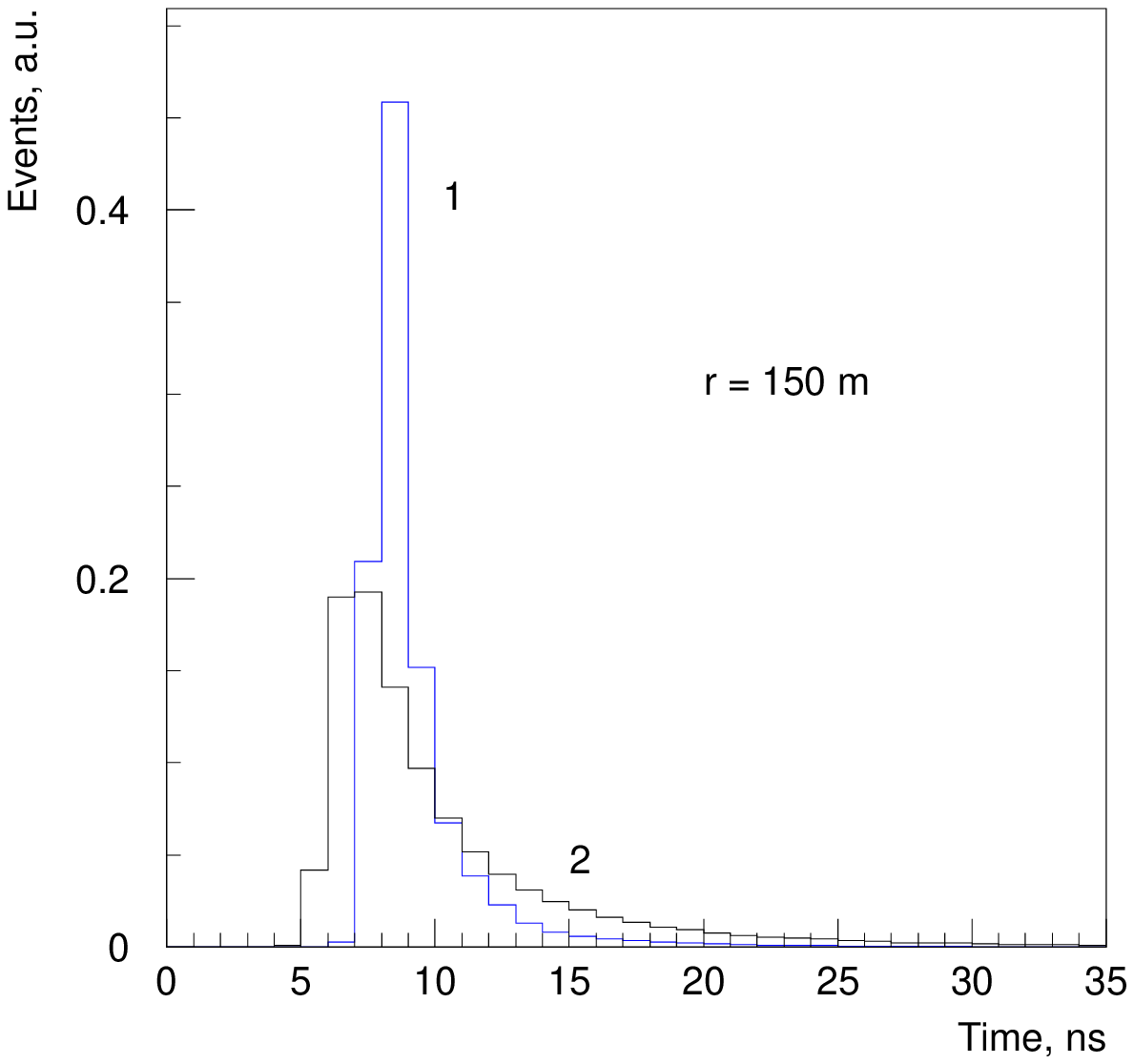}
\caption{Time profile of \v{C}erenkov light flashes simulated for
impact distance of 150~m and two observational levels of 2 (1) and
5 (2) km above the sea. Primary energy of simulated air-showers is
10~GeV. Label a.u. along Y-axis stays for the arbitrary units.
\label{fig:3}}
\end{figure}

Shape of time pulse can be characterized by a few parameters, e.g.
$t_1=t_{30\%}-t_{10\%}$, $t_2=t_{50\%}-t_{10\%}$,
$t_3=t_{90\%}-t_{10\%}$, where $t_{10\%}$, $t_{30\%}$, $t_{50\%}$,
and $t_{90\%}$ give the time tags, which are defined as, e.g.
$\int_o^{t_{10\%}}p(t)dt=0.1$. Results of calculations are shown in
Figure~\ref{fig:2}. One can see that for a 500~$\rm [gr/cm^2]$
observational level time pulses are substantially broader at impact
distances beyond 100~m. Hereafter we are dealing with \v{C}erenkov
light images averaged over a sample of simulated 10~GeV gamma-ray
showers. For an impact distance of ca. 250~m, the time pulse of
\v{C}erenkov light flash recorded at high altitude in the atmosphere
will be a factor of 2 broader than for the same impact distance at a
conventional altitude of 2~km above sea level (see
Figure~\ref{fig:3}). Integration over the time pulse yields a total
number of \v{C}erenkov photons in a flash. Therefore, for a given
flux of night sky background light, {\it a signal-to-background
ratio might be correspondingly lower by factor of 2 for a high
altitude site}. For high energy gamma-ray showers ($E_o\geq$100~GeV)
\v{C}erenkov pulses recorded at 5~km above sea level might be as
long as 50~ns. Registration of these pulses will occur in the regime
highly dominated by night sky background.

It is worth noting that, at the time of writing, there is no well
established altitude dependence of a flux of night sky background
light available. In general this parameter is considered to be very
specific for each individual observational site. Apparently the high
altitude sites provide substantially reduced attenuation and
consequently more starlight from the individual stars, which is in
fact a background for the \v{C}erenkov telescopes. However, the
effect of bright stars might be diminished simply by switching off
the high voltage for those camera pixels (PMTs) collecting direct
star light. Such procedure usually runs in automatic mode while
taking the observational runs. At the same time one might expect a
significant increase in flux of background light photons within the
UV wavelength range (200-300~nm). However, conventional imaging
cameras are not very sensitive in a wavelength range well below
300~nm.

Even though there are good reasons to believe that this flux will,
in fact, be much lower (on average) for high altitude sites, it
still needs to be measured at any chosen observational site.

For the sake of thoroughness one should mention that the high
altitude sites will noticeable increase the probability that the
ionizing particles, such as atmospheric electrons and muons of low
energy cosmic rays, can directly hit the camera PMTs. It will lead
to some random increase in background light over the camera pixels.
However, dedicated calculations are needed in order to quantify this
additional component of the background, which fall unfortunately out
of the area of this paper.

\subsection{Images}
A two-dimensional distribution of the \v{C}erenkov light intensity
in the telescope's focal plane (image), $q(\vec{\theta})$, can be
effectively used to derive detailed information about shower
orientation and shape. Phenomenology of \v{C}erenkov light images
was discussed in \cite{hillas}. For an ellipsoid-like image, the
orientation of its major axis constrains the shower orientation in
space with respect to the telescope optical axis. Images recorded at
relatively small impact distances from the shower axis (R$\le$100~m)
have circular shape, and an accurate determination of the major axis
is quite difficult. For impact distances beyond 100~m the image
ellipsoid has a well defined shape, and the ratio of its angular
size measured along a major axis to the corresponding angular size
of the image measured along a minor axis is ca. 2 and above. At the
same time at large impact distances (R$\gg$200~m) the total number
of \v{C}erenkov photons in an image becomes rather low and high
fluctuations prevent accurate measurement of the image orientation.
Those two effects finally constrain the range of optimum impact
distances for effective shower reconstruction. As mentioned above,
the advantage of the high altitude site is mainly associated with an
enhancement in \v{C}erenkov light density at small distances to the
shower axis (R$\leq$100~m). However, in this range of impact
distances \v{C}erenkov light images tend to have poorly determined
orientation.

\begin{figure}[t]
\centering
\includegraphics[height=5cm]{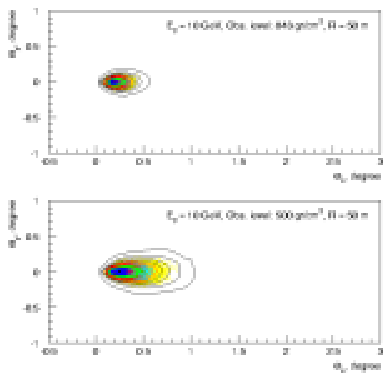}\hspace*{10mm}
\includegraphics[height=5cm]{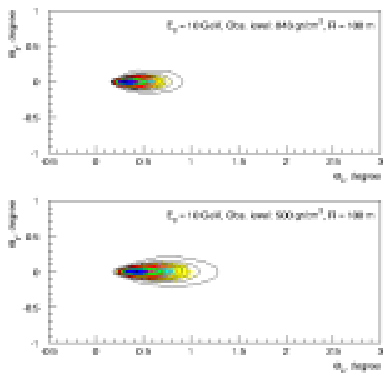}\\ \vspace*{10mm}
\includegraphics[height=5cm]{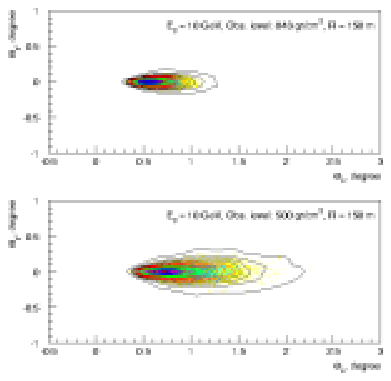}\hspace*{10mm}
\includegraphics[height=5cm]{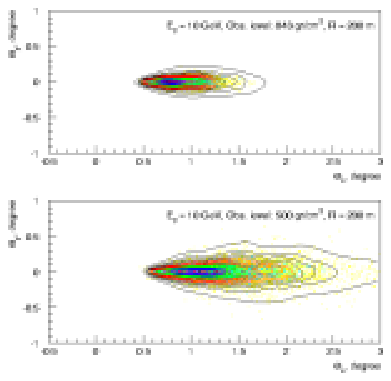}
\caption{Images of \v{C}erenkov light calculated at different
altitudes above sea level for a number of impact distances of the
telescope to the shower axis as indicated in the pictures. Images
were averaged over a sample of 10~GeV gamma-ray showers. The
contour plots were drawn starting from the position of maximum
intensity with the isoline increment of $ln2$.   \label{fig:4}}
\end{figure}

Average \v{C}erenkov light images for two observational heights of
2 and 5~km, respectively, are shown in Figure~\ref{fig:4}. Note
that scales used along X- and Y-axis are not identical. Detailed
comparison of those images revealed two major differences in their
shape.

\begin{enumerate}
\item Images of 10~GeV gamma-ray showers recorded at high altitude
above sea level are substantially larger in size (see
Figure~\ref{fig:4} and Table~\ref{tab:2}). This is mainly because
showers are located at a relatively smaller distance to the
telescope, than for a conventional observational site of 2~km above
the sea. Calculations show that the angular size of an image
measured along the major axis, the image length, increases
considerably faster with altitude than the angular size of image
measured along a minor axis, the image width. This can be easily
understood by comparing the ratio of the actual scale of a shower
longitudinal development over the distance of shower maximum to the
telescope, which is located at two altitudes of 2 and 5~km above the
sea, respectively. In a simplified toy model the length of the image
will scale with height of observation level above sea level, $H_o$,
as
\begin{equation} L\propto (H_{max}-H_o),~
R<<(H_{max}-H_o).\end{equation}  \item Images recorded at high
altitude must have considerably larger displacement from the
center of the focal plane. Coming closer to the shower maximum
($H_{max}$ is the height of the shower maximum), the shower will
be apparently seen in \v{C}erenkov light at larger angle
\begin{equation}
\Theta \propto \tan^{-1}(\frac{R}{H_{max}-H_o}) \end{equation}
with respect to the optical axis.
% (see also \cite{memo}).
\end{enumerate}

\begin{table}[t]
\centering \caption{Area $A,~[deg^2]$ and effective size,
$r_o=\sqrt{ab}~[deg]$, (in this Table both values are given in a
format of $A/r_o$) of \v{C}erenkov light images calculated for two
observational levels of 2 and 5~ km, and for a few impact
distances.\label{tab:2}}
\begin{tabular}{llcccc}
\hline\noalign{\smallskip}
H [km] & D [$\rm gr/cm^2$] & R [m] = 50 & 100 & 150 & 200 \\
\noalign{\smallskip}\hline\noalign{\smallskip}
2.2 & 843 & 0.16/0.22 & 0.24/0.27 & 0.31/0.32 & 0.44/0.37 \\
5   & 500 & 0.40/0.36 & 0.54/0.41 & 0.87/0.53 & 1.27/0.64 \\
\noalign{\smallskip}\hline
\end{tabular}
\end{table}

One can see in Figure~\ref{fig:1} that the density of \v{C}erenkov
photons from a 10~GeV gamma-ray shower is approximately the same at
both observational heights for impact distances above 150~m from the
shower axis. Thus, at large impact distances, the difference in
image shape mentioned above is of purely geometrical origin, which
is independent of the image size (the total number of photons in the
image). At first glance, large images at high altitudes might offer
better resolution for a camera of crude pixellation. However, for a
10~GeV gamma-ray shower images will always be significantly affected
by contamination of background light and reduction of background
light per camera pixel. Large images recorded at high altitudes
yield considerably lower \v{C}erenkov light density per 1~$str$ for
fixed image's size. {\it For the flux of night sky photons, as
measured at conventional observational level of 2~km above the sea,
the image of a 10~GeV gamma-ray shower recorded at high altitudes
will have higher contamination of background photons per camera
pixel of any size}.

As mentioned above, the images recorded at high altitudes must have
large displacements from the camera center. This issue stringently
constrains the design of the camera for the telescope placed at high
altitudes, in particularly the field of view has to be larger. {\it
Cameras of a narrow field of view will be drastically limited in
their ability to detect gamma-rays at high energies}.

\begin{figure}[htbp]
\centering (a) \includegraphics[height=3cm]{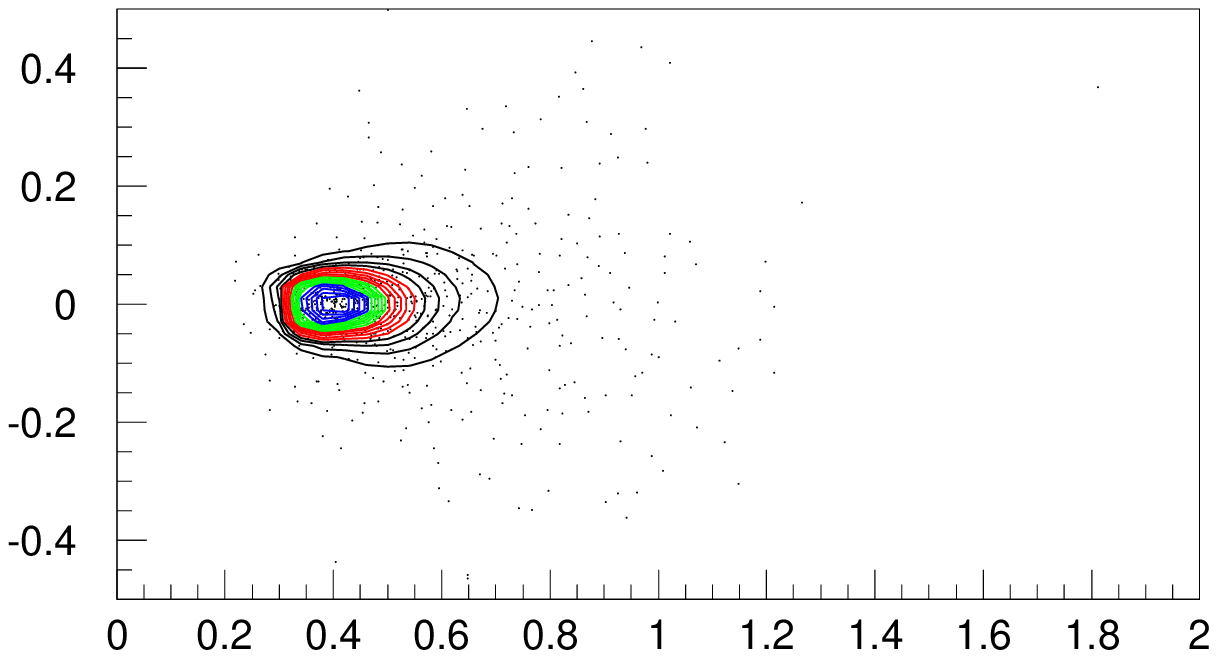} \\
\vspace*{2mm} (b) \includegraphics[height=3cm]{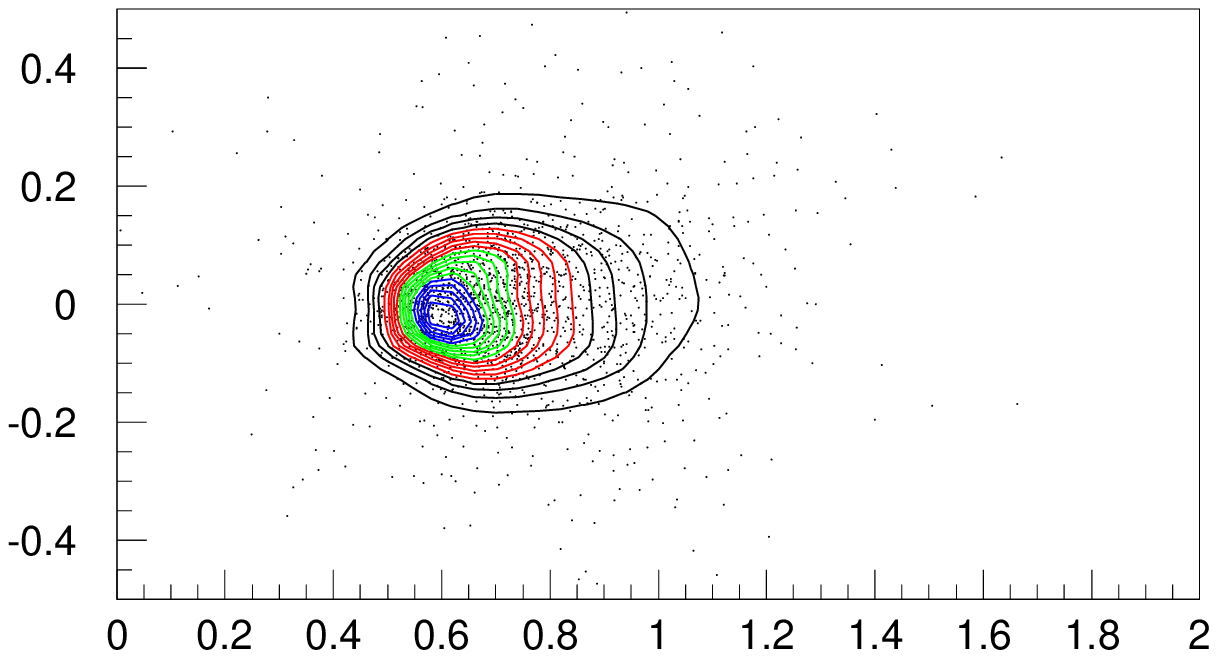}\\
\vspace*{2mm} (c) \includegraphics[height=3cm]{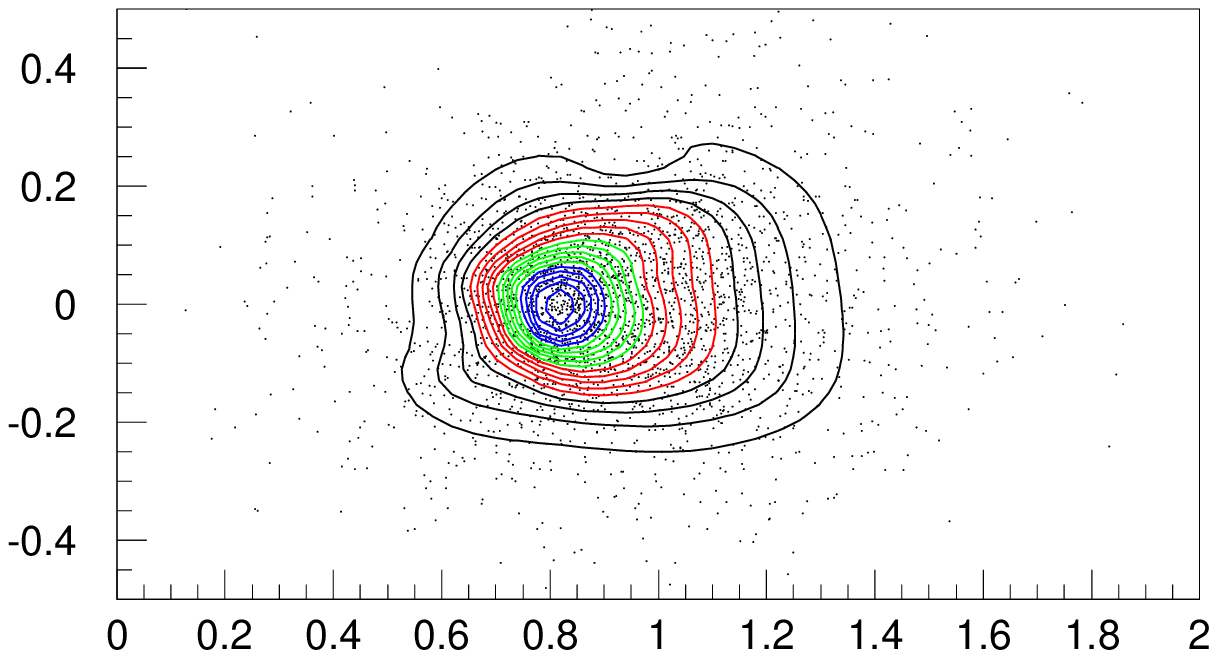}\\
\vspace*{2mm} (d)
\includegraphics[height=3cm]{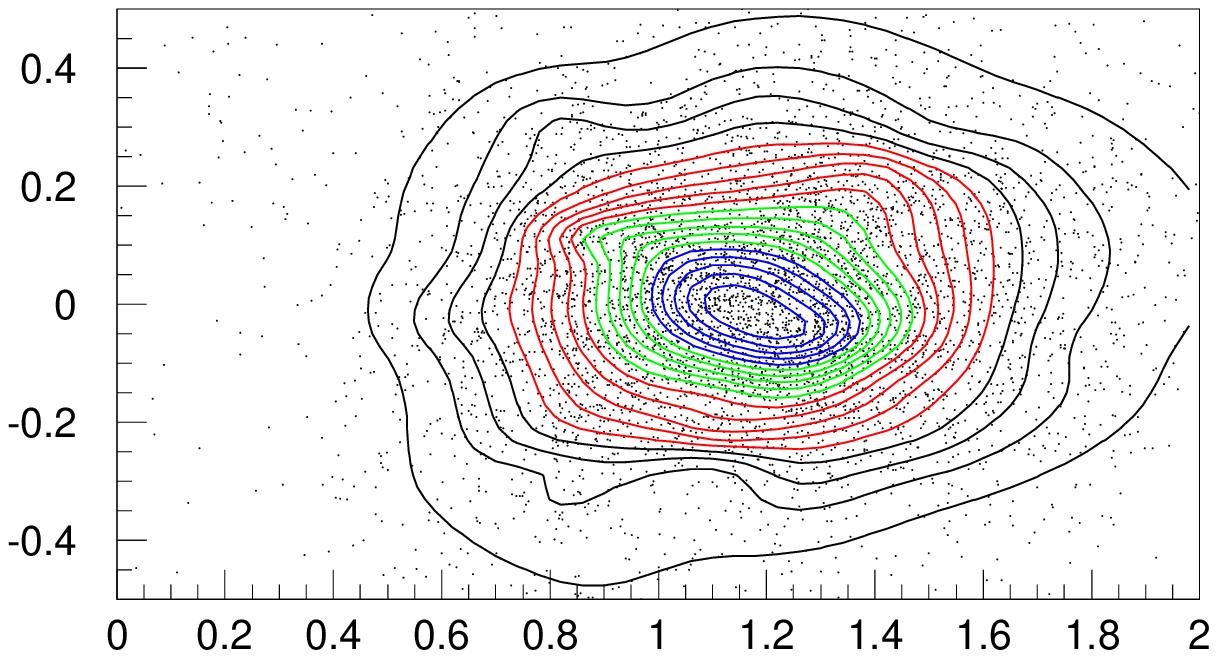}\\\vspace*{10mm}
\hspace*{4mm} \includegraphics[height=3cm]{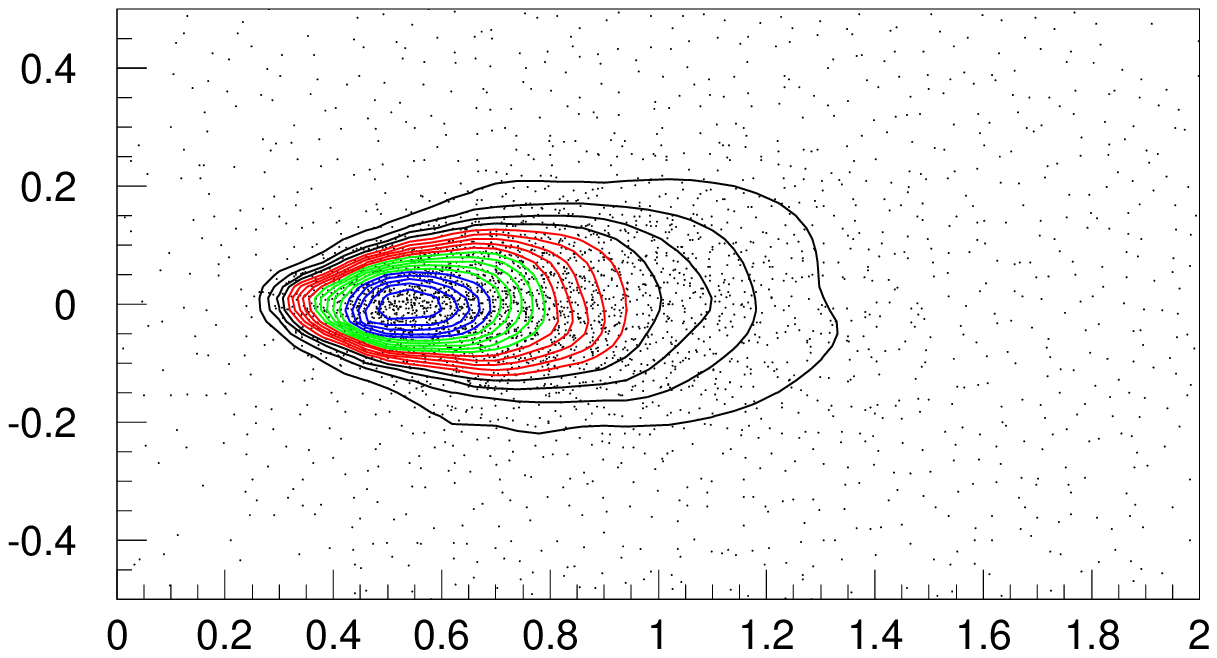} \caption{Average
images of \v{C}erenkov light from a 10~GeV gamma-ray shower,
calculated at an observation level of 2~km above sea level, and
for an impact distance of 150~m from the shower axis. An
additional selection on arrival time of \v{C}erenkov  photons was
applied. Corresponding time windows were 7.25-7.5~ns (a);
8.25-8.5~ns (b); 9-9.5~ns (c); 10-12~ns (d). Each of (a)-(d) plots
contains approximately the same number of photoelectrons. The
image shown in the lower panel was generated without time
selection. \label{fig:5}}
\end{figure}

\subsection{Time-dependent imaging}
One can try to suppress contamination of night sky background light
in an image by using a very narrow time gate. This may be tuned
exactly, for example, to a rising edge, maximum, or tail of a time
pulse. However, it leads to a trade off between losing a substantial
fraction of \v{C}erenkov  photons and on the other hand a severe
reduction of night sky background. This approach is illustrated by
the images shown in Figure~\ref{fig:5}. Photons emitted at the very
beginning of the shower development in the atmosphere, which are
mainly arriving at the front of the time pulse of a flash, form the
so-called image "conk"\footnote[3]{Non-standard definition of a
strongly elongated part of the comet like image, which has a
relatively high photoelectron density.}. The image of shower
electrons, which propagate further into the atmosphere, shifts
further away from the center of telescope's focal plane. Multiple
scattering of low energy cascade electrons at later stages of the
shower development becomes very important and it results in a very
broad image as shown in Figure~\ref{fig:5}. One can see also in
Figure~\ref{fig:5} that \v{C}erenkov light {\it images for such
narrow time windows are in fact rather symmetric in shape, and they
are not so good for reconstruction of image orientation}. At the
same time, an image which contains all registered \v{C}erenkov
photons ($\rm 0~ns~<~t~<~14~ns$), has a regular ellipsoid-like
shape. Note that photons, which are significantly delayed with
respect to the front of time pulse, will be hitting the outer edge
of the image, whereas background light photons are apparently
dominant.

Time-dependent imaging seems to be a rather promising approach in
improvement of gamma/hadron separation of extremely low energy
events. Despite that {\it present analysis does not reveal an
evident improvement using time-dependent imaging}, it could be
perhaps very effective in analysis based on centroid positions in a
few triggered telescopes in an array.

\section{Conclusions}

In this paper we have attempted to perform a comparative analysis of
\v{C}erenkov light images simulated for two observational levels of
2 and 5~km above sea level, respectively. We used as the basis of
our calculations a 30~m diameter telescope, which is a conceivable
size for a future \v{C}erenkov telescope project. The system aspect
was not discussed here, but all conclusions are obviously relevant
to any telescope of a possible future array. One has to mention that
for a future detector, approaching a very low energy threshold of
about 10~GeV or even below, a contamination of night sky background
in the registered shower images is in fact a very important issue.
It might finally constrain the choice of the observational site for
such low threshold arrays of \v{C}erenkov telescopes.

Results reported here generally confirm that the observational site
at higher altitude provides substantially higher \v{C}erenkov light
density. However this enhancement occurs only within the area
limited by roughly 100~m from shower axis. Therefore all recorded
gamma-ray showers well below 10~GeV (here we assume that the
telescope has in fact sufficient area of the reflector, see
discussion in Section~\ref{ss:1}) will concentrate in a region of
relatively small impact distances. Those images are not so clearly
elongated in shape, which makes reconstruction of the image's
orientation rather difficult.

Flashes of \v{C}erenkov light from 10~GeV gamma-ray showers have
broader time pulses for the impacts beyond 100~m at higher altitude.
Corresponding images are considerably broader as well. Both of these
two effects might apparently substantially increase the background
light contamination in an image. Images recorded at high altitude
must be further displaced from the center of telescope's focal
plane, than for conventional altitude of ca. 2~km above sea level.
They also, most probably, require larger field of view.

The time-depending imaging is a very promising approach in further
development of advanced analysis for observation of low energy
gamma-ray showers, but it might be not very effective in resolving
the problem of dominating night sky background light in the
recorded images of such low energy gamma-rays.

One can briefly conclude that an observational site at high
altitude might provide {\it further modest reduction of the energy
threshold} of a future detector, even though the shape of time
pulses and in particular the topology of the two-dimensional
angular distribution of \v{C}erenkov light flashes recorded at
extremely high altitudes are palpably {\it less preferable} for
imaging of gamma-ray showers above 10~GeV.

We hope that the results presented here may help to increase the
understanding of changes in topology of \v{C}erenkov light images
after an increase in the observational level from its conventional
height of ca. 2~km up to 5~km above sea level.

Ultimately, the choice of a site for the next generation of
ground-based imaging atmospheric \v{C}erenkov detector, which is
widely believed to be a system of 30~m class telescopes, will depend
on both scientific and political issues relating to funding,
international collaboration etc. The move to lower energy threshold
is likely to remain a significant drive for the science. However,
one has to consider all trade-offs, i.e. reasonable altitude, low
level of night sky background, need for the robotic telescopes etc,
in selecting candidate sites for such a detector to optimize the
scientific goals.

\section*{Acknowledgements} I would like to thank the referees, who
remain anonymous, for the invaluable comments and suggestions, which
have improved the quality of this paper.

\end{document}